\pdfoutput=1
\documentclass[twocolumn,aps,showpacs,nofootinbib,superscriptaddress,prd]{revtex4-1}
\usepackage{mathrsfs}
\usepackage{epsfig}
\usepackage{textcomp}
\usepackage{amsmath}
\usepackage{graphicx}
\usepackage{amssymb}
\usepackage{epstopdf}
\usepackage{slashed}
\usepackage{mathtools}
\usepackage{color}
\allowdisplaybreaks[4]
\newcommand{\sect}[1]{{\bf{#1.} }}

\begin{document}

\title{Direct reduction of multiloop multiscale scattering amplitudes}

\author{Yefan Wang}
\email{wangyefan@ihep.ac.cn}
\affiliation{Institute of High Energy Physics, Chinese Academy of Sciences, Beijing
100049, China}
\affiliation{School of Physics Sciences, University of Chinese Academy of Sciences,
Beijing 100039, China}

\author{Zhao Li}
\email{zhaoli@ihep.ac.cn}
\affiliation{Institute of High Energy Physics, Chinese Academy of Sciences, Beijing
100049, China}
\affiliation{School of Physics Sciences, University of Chinese Academy of Sciences,
Beijing 100039, China}

\author{Najam ul Basat}
\email{najam@ihep.ac.cn}
\affiliation{Institute of High Energy Physics, Chinese Academy of Sciences, Beijing
100049, China}
\affiliation{School of Physics Sciences, University of Chinese Academy of Sciences,
Beijing 100039, China}

\begin{abstract}
We propose an alternative approach based on series representation to directly reduce multi-loop multi-scale scattering amplitude into set of freely chosen master integrals. And this approach avoid complicated calculations of inverse matrix and dimension shift for tensor reduction calculation. During this procedure we further utilize the Feynman parameterization to calculate the coefficients of series representation and obtain the form factors. Conventional methodologies are used only for scalar vacuum bubble integrals to finalize the result in series representation form. Finally, we elaborate our approach by presenting the reduction of a typical two-loop amplitude for W boson production.
\end{abstract}

\maketitle

\sect{Introduction}
The CERN Large Hadron Collider (LHC) is the most accurate experiment on the
elementary particle physics at present,
and the next generation lepton colliders have been proposed aiming
at higher accuracy.
They all demand the high precision theoretical predictions to include higher orders of
either electroweak or QCD corrections \cite{Bendavid:2018nar}.
However, the higher order corrections may become seriously challenging due to the
evaluation of the multi-loop Feynman diagrams, which usually can be decomposed into several steps of calculations.
And practically one of the most difficult calculations
is to reduce the loop amplitude into linear combination of master integrals.

For the one-loop amplitude many different reduction algorithms have been developed
after decades of effort.
The Passarino-Veltman reduction
algorithm \cite{Passarino:1978jh,Consoli:1979xw,Veltman:1980fk,Green:1980bd}
has been widely used in enormous number of investigations
on the next-to-leading order (NLO) effects for the Standard Model (SM) processes
and some new physics processes.
Later the implementation of unitarity
algorithm
\cite{Bern:1994zx,Bern:1994cg,Britto:2004nc,Roiban:2004ix,Ossola:2007bb,Forde:2007mi,Giele:2008ve,Ellis:2008ir}
on the one-loop amplitude provided very interesting and inspiring prospect on the
amplitude structure.
Meanwhile the algorithm based on unitarity also presents excellent numerical
efficiency.
Consequently, by implementing these modern reduction algorithms the SM NLO
calculations have been
automated
\cite{Berger:2008sj,Bevilacqua:2011xh,Cascioli:2011va,Badger:2012pg,Cullen:2014yla,Alwall:2011uj,Actis:2016mpe}. Other methods can be found from \cite{Buccioni:2018zuy,Buccioni:2017yxi,delAguila:2004nf,Bern:2005cq,Denner:2005nn,vanHameren:2009vq}.

At the multi-loop level in the consideration of efficiency the amplitude has to be reduced into linear
combination of finite number of master integrals \cite{Smirnov:2010hn},
which can be further calculated analytically or numerically.
In contrast to the one-loop case,
the achievement of multi-loop reduction conventionally includes two separate steps,
i.e. the tensor reduction and the scalar integral reduction using integration by part (IBP) identities \cite{Laporta:2001dd}.

First the tensor reduction is used to isolate the loop momenta from fermion chains, polarization vectors or product of them,
which will be factorized out of the loop integral to construct the form factors.
Specifically one of conventional approaches is the projection method
\cite{Binoth:2002xg,Glover:2004si}
that has been commonly used in the calculations of high order QCD corrections to
the Higgs
production \cite{Gehrmann:2011aa,Melnikov:2017pgf,Boggia:2017hyq}
and the vector boson productions \cite{Gehrmann:2011ab,Gehrmann:2015ora}.
The key to projection method is the projector basis relying on the analytic
inversion of projection matrix.
However, for some complicated scattering processes,
e.g. the full next-to-next-to-leading order QCD correction
to single-top production \cite{Assadsolimani:2014oga},
the project matrix could become so big that its inversion may seriously challenge
the computation resource.
Another approach for tensor reduction is Tarasov's
method \cite{Tarasov:1996br} based on Schwinger parameterization
\cite{Speer:1974cz,Bergere1974}.
It can avoid irreducible numerator but shift the space-time dimensions of
obtained scalar integrals.
Thus it is inevitable to shift the dimensions of scalar integrals back
to the conventional $D$-dimension or the same dimension at least.
And this commonly needs to resolve the dimension recurrence relations,
which however is as difficult as the matrix inversion in projection method.
Besides another popular approach is using IBP identities
\cite{Tkachov:1981wb,Chetyrkin:1981qh},
which however also confronts serious difficulties in the multi-scale processes.
During this modern age of evaluation, computational algebraic based
algorithms \cite{Mastrolia:2011pr,Badger:2012dp,Zhang:2012ce} successfully
implemented on N=4 Yang-Mills theory and numerical unitarity method
for multi-gluon amplitudes \cite{Abreu:2017xsl,Abreu:2017hqn,Badger:2017jhb,Abreu:2018zmy,Abreu:2018jgq}.

Then after the successful tensor reduction the loop amplitude becomes linear
combination of
scalar integrals, whose number could be order ${\mathcal O}(10^4)$ for complicated
processes.
Consequently as the second step usually the IBP reduction
is introduced to reduce the scalar integrals into
a much smaller number of master integrals.
The most popular method for IBP reduction is Laporta
algorithm \cite{Laporta:2001dd}, which has been implemented by several codes
\cite{Chetyrkin:1981qh,vonManteuffel:2012np,Lee:2013mka,Smirnov:2014hma,Georgoudis:2016wff,Maierhoefer:2017hyi,Smirnov:2019qkx}.
Another interesting method \cite{Georgoudis:2016wff,Georgoudis:2017iza}
for IBP reduction recently has been developed based on algebraic geometry.
Due to the fact that IBP reduction relies heavily on the IBP relations
the choice of master integral set cannot be arbitrary,
so the resulting reduction expressions may
confront unacceptable inflation \cite{Borowka:2016ehy,Jones:2018hbb}.
Therefore, to efficiently evaluate the multi-scale multi-loop amplitude
one better keep the freedom to choose master integrals.
And this can be achieved by series representation \cite{Liu:2018dmc},
which in fact can also be used to solve the tensor reduction as we will show in the
following.

In this paper, based on the series representation \cite{Liu:2017jxz,Liu:2018dmc},
we propose an alternative reduction approach that can directly reduce loop amplitude to
master integrals
so that the complexity of tensor reduction and IBP reduction can be relieved.
In next section the main idea will be explained in detail.
Then its application on one typical two-loop diagram
of W boson production as an example will be shown.
Finally the conclusion is made.

\sect{Amplitude Reduction}
Recently, series representation of Feynman integral
has been proposed
to reduce the scalar integrals into master integrals \cite{Liu:2018dmc}
and
to numerically evaluate the master integrals \cite{Liu:2017jxz}.
It is very promising since it can be applied to multi-scale multi-leg integrals
and has freedom to choose master integrals.
Intriguingly we find that the series representation
can also be directly implemented on the loop amplitude,
which in general can be written as
\begin{equation}
{\mathcal M} = \int
{\mathbb D}^L q
\frac{N(\{q_j\}_{j=1}^{L},\{k_e\}_{e=1}^E)}
{\prod_{i=1}^n \mathcal{D}_i^{\nu_i} },
\end{equation}
where
${\mathbb D}^L q \equiv \prod_{\ell=1}^L {\mathrm d}^{D}q_\ell$.
$\{q_j\}_{j=1}^{L}$ are $L$ loop momenta, $\{k_e\}_{e=1}^E$ are $E$ external
momenta
and $\{\mathcal{D}_i\}_{i=1}^n$ are the denominators of loop propagators.
Numerator $N(\{q_j\}_{j=1}^{L},\{k_e\}_{e=1}^E)$ may contain fermion chains, polarization
vectors or both.

In order to obtain the expression of loop amplitude in series representation,
we first modify all the denominators,
\begin{equation}\label{modify}
\frac{1}{{\mathcal D}_i}\equiv \frac{1}{P_i^2-m_i^2}  \rightarrow
\frac{1}{\widetilde {\mathcal D}_i} \equiv \frac{1}{P_i^2-m_i^2+\imath\eta},
\end{equation}
where $P_i\equiv Q_i+K_i$ is the momentum of the $i$-th propagator. $Q_i$ and $K_i$
are defined as linear combinations of loop momenta and external momenta, respectively.
Therefore we obtain the modified amplitude $\widetilde {\mathcal M}(\eta)$,
which depends on auxiliary parameter $\eta$. The mass dimension of $\eta$ is same as the mass dimension of $m_i^2$. With the help of the parameter $\eta$, any modified amplitude can be defined as series representation. After the reduction the physical original amplitude can be obtained in the limit of $\eta\rightarrow 0^+$ 
\begin{eqnarray}
{\mathcal M}
= \lim_{\eta\to 0^+} \widetilde {\mathcal M}(\eta).
\end{eqnarray} 

The modified loop amplitude can be decomposed as linear combination of tensor integrals
\begin{eqnarray}
\widetilde {\mathcal M}(\eta) =
\sum_{\substack{\mu_1\dots\mu_R\\ \ell_1\dots\ell_R}}
N_{\mu_1\dots\mu_R,\ell_1\dots\ell_R}(\{k_e\}_{e=1}^E)\widetilde G^{\mu_1\dots\mu_R}_{\ell_1\dots\ell_R},
\end{eqnarray}
where
$N_{\mu_1\dots\mu_R, \ell_1\dots\ell_R}(\{k_e\}_{e=1}^E)$ is the coefficient of tensor integral.
\begin{eqnarray}
\widetilde G^{\mu_1\dots\mu_R}_{\ell_1\dots\ell_R} \equiv
\int
{\mathbb D}^L q~
\frac{
q^{\mu_1}_{\ell_1}\dots q^{\mu_R}_{\ell_R}
}
{
\prod_{i=1}^n
[(Q_i+K_i)^2-m_i^2+\imath\eta]^{\nu_i}
}.
\end{eqnarray}
The summation is over all tensor structures in the given amplitude.
By using Feynman parameterization \cite{Heinrich:2008si} for tensor integrals,
we can express the tensor integral as
\begin{eqnarray}\label{tensor}
\widetilde G^{\mu_1\dots\mu_R}_{\ell_1\dots\ell_R}
&&=\frac{(-1)^{N_\nu}}{\prod_{j=1}^n\Gamma(\nu_j)}
\int \prod_{j=1}^n {\mathrm d}x_j~ x_j^{\nu_j-1}
\delta(1-\sum_{l=1}^n x_l)
\nonumber
\\ &&
\times \sum_{m=0}^{\left[R/2\right]}
\frac{ \Gamma(N_\nu^{(m)}) }{ (-2)^m }
\left[ (\tilde M ^{-1} \otimes g)^{(m)}\widetilde \ell^{(R-2m)}\right]^{\Gamma_1,\dots,\Gamma_R}
\nonumber
\\&&
\times
U^{-D/2+m-R} \left(\frac{F}{U}-\imath \eta\right)^{-N_\nu^{(m)}},
\end{eqnarray}
where $N_\nu \equiv \sum_{i=1}^n \nu_i$ and $N_\nu^{(m)}\equiv N_\nu -m -LD/2$.
$U$ and $F$ are the first and second Symanzik polynomials, respectively. Here $U$ and $F$ are polynomials of Feynman parameters $\{x_i\}$, and $F$ can also include the 
masses and the scalar products of the external momenta. $\tilde M$ is the matrix of $\{x_i\}$ and $\widetilde \ell$ depends on the external momenta. $m$ is defined as "metric rank" to indicate the number of metric tensor generated in each term of the summation in Eq. \eqref{tensor}. The explicit definitions of symbols in the square bracket can be found in Ref. \cite{Heinrich:2008si}. 
An important observation is that auxiliary parameter $\eta$ only appears in the last bracket.
By using Taylor series for $\eta\to\infty$ one can obtain
\begin{eqnarray}
\left(\frac{F}{U}-\imath \eta\right)^{-N_\nu^{(m)}} =&& (-\imath\eta)^{-N_\nu^{(m)}}\sum_{p=0}^\infty \left(
\begin{matrix}
-N_\nu^{(m)}
\\
p
\end{matrix}
\right)\nonumber\\&&\times\frac{F^p}{U^p(-\imath\eta)^p}.
\end{eqnarray}
Now it can be seen that the exponent $p$ in $F^p$ is a non-negative number, so that 
the difficulty of dealing with fraction polynomial $F^{-N_\nu^{(m)}}$ can be avoided. 
By using direct expansion $F^p$ can be expressed as the polynomial of $\{x_i\}$, 
then all the tensor structures are 
only related to the external momenta. Consequently the external momenta 
can be attached to fermion chains or the polarization vectors to generate the form factors. 
And the coefficients of form factors become integrals on Feynman parameters $\{x_i\}$, for instance
\begin{eqnarray}
\int \prod_{j=1}^n {\mathrm d}x_j~ x_j^{n_j-1} \delta(1-\sum_{l=1}^n x_l) U^{-\widetilde D/2},
\end{eqnarray}
where $\widetilde D$ can be different from the space-time dimension $D$. We can define an equivalence relation $\sim$ between Feynman parameter indices, such that
\begin{eqnarray}
i \sim j \quad \text{if} \lim_{\substack{K_i \to 0\\m_i \to 0}}\widetilde{\mathcal D}_i= \lim_{\substack{K_j \to 0\\m_j \to 0}}\widetilde{\mathcal D}_j.
\end{eqnarray}
Then we can divide the Feynman parameters index set $\{i\}^n_{i=1}$ into $X = 2^L-1$ equivalence classes $[{i_1}],\dots,[{i_X}]$.
For each equivalence class, we can insert one unit integral, e.g.,
\begin{eqnarray}
\int {\mathrm d} y_1 \delta\left(y_1-\sum_{j \in [{i_1}]} x_{j}\right)=1.
\end{eqnarray}
Meanwhile because $U$ can be constructed from the 1-tree cut on the Feynman loop diagram \cite{Bogner:2010kv},
it can be found that $U$ only depends on $\{y_i\}$.
Then the parameters $\{x_{j}\}$ can be integrated along with the inserted $\delta$-functions as
\begin{eqnarray}
&&\int \prod_{k=1}^n {\mathrm d}x_k~ x_k^{n_k-1} \delta\left(1-\sum_{l=1}^n x_l
\right) U^{-\widetilde D/2} \nonumber\\
=\label{key}&&\int \prod_{m=1}^X\left( \prod_{j\in[i_m]}\left({\mathrm d}x_j~ x_j^{n_j-1}\right)\right) \delta\left(1-\sum_{l=1}^n x_l\right) U^{-\widetilde D/2} \nonumber\\
=&&\int \prod_{m=1}^X\left( \prod_{j\in[i_m]}\left({\mathrm d}x_j~ x_j^{n_j-1}\right){\mathrm d}y_m~ \delta\left(y_m - \sum_{p \in [i_m]}x_p\right)\right) \nonumber\\
&&\times\delta\left(1-\sum_{l=1}^n x_l\right) U^{-\widetilde D/2}\nonumber\\
=&&\int \prod_{m=1}^X\left({\mathrm d}y_m~\frac{\prod_{j \in [i_m]}\Gamma(n_j)}{\Gamma\left(\sum_{j \in [i_m]} n_j\right)} y_m^{\left(\sum_{j \in [i_m]} n_j\right)-1}\right)\nonumber\\
&&\times\delta\left(1-\sum_{l=1}^X y_l\right)U^{-\widetilde D/2}.
\end{eqnarray}
And finally the integrals on $\{y_i\}$ can be reconstructed as vacuum bubble integrals, for instance at two-loop level
\begin{eqnarray}
I_{\nu_1,\nu_2,\nu_3}^{(vac),\widetilde D} &\equiv& \int \frac{{\mathrm d}^{\widetilde D} q_1 {\mathrm d}^{\widetilde D} q_2}
{ [q_1^2+\imath]^{\nu_1}[q_2^2+\imath]^{\nu_2}[(q_1+q_2)^2+\imath]^{\nu_3} }
\nonumber \\
&=&
(-i)^{\widetilde D+N_\nu} \int
\frac{y_1^{\nu_1-1} {\mathrm d}y_1}{\Gamma(\nu_1)}
\frac{y_2^{\nu_2-1} {\mathrm d}y_2}{\Gamma(\nu_2)}
\frac{y_3^{\nu_3-1} {\mathrm d}y_3}{\Gamma(\nu_3)}
\nonumber \\ && \times
\delta(1-y_1-y_2-y_3)\Gamma(N_\nu- \widetilde D) U^{-\widetilde D/2}.
\end{eqnarray}
For the remaining scalar vacuum bubble integrals, we can further implement the IBP reduction \cite{Chetyrkin:1981qh} to reduce $I_{\nu_1,\nu_2,\nu_3}^{(vac),\widetilde D}$ to $I_{1,1,1}^{(vac),\widetilde D}$ and $I_{1,1,0}^{(vac),\widetilde D}$. 
Then we can implement the dimension shift operation to reduce them to $I_{1,1,1}^{(vac),D} $ and $I_{1,1,0}^{(vac),D}$.
Finally the modified loop amplitude can be expressed as the series representation
in terms of vacuum bubble master integrals in $D$ dimension.
It is necessary to emphasize that the IBP reduction and the dimension shift operation
are implemented only on the vacuum bubble integrals,
which are process independent and can be easily prepared once for all.

Obviously now we have successfully achieved the tensor reduction for loop amplitude.
Finally we can rewrite the modified loop amplitude as
\begin{equation}
\widetilde {\mathcal M}(\eta) = \sum_i {\mathcal C}_i {\mathcal F}_i,
\end{equation}
and
\begin{equation}\label{e1}
{\mathcal C}_i = \eta^{L{D}/2-N_\nu+ m_i^{\rm max}}
\sum_{p=0}^\infty \sum_j {\mathcal A}_{0pj}\eta^{-p}I^{(vac),D}_{L,j},
\end{equation}
where ${\mathcal F}_i$ is the form factor and ${\mathcal C}_i$ is the relevant
coefficient.
$m_i^{\rm max}$ is the maximum of the metric ranks of the terms that contribute to ${\mathcal F}_i$.
And $I^{(vac),D}_{L,j}$ represents the $j$-th $L$-loop vacuum bubble master
integral. The series coefficient ${\mathcal A}_{0pj}$ only explicitly depends on linear independent kinematic variables $\{s_1,\dots,s_t\}$ and space-time dimension $D$. Since in the following we will focus on one of the coefficients ${\mathcal C}_i$ to demonstrate the reduction procedure, for simplicity of the formula, the index $i$ dependence for ${\mathcal A}_{0pj}$ is suppressed. Here we define tuple $s \equiv (s_1,\dots,s_t)$ and monomial
\begin{eqnarray}
s^{\alpha} \equiv s_1^{\alpha_1}\cdots s_t^{\alpha_t},
\end{eqnarray}
where $\alpha = (\alpha_1,\dots,\alpha_t)$ is a $t$-tuple of nonnegative integers. $\left|\alpha\right| = \alpha_1+\cdots+\alpha_t$ is the total degree of monomial $s^{\alpha}$. Then ${\mathcal A}_{0pj}$ can be written as
\begin{eqnarray}
{\mathcal A}_{0pj} = \sum_{\substack{{\alpha_1,\dots,\alpha_t}\\|\alpha| = \mathrm{dim}({\mathcal A}_{0pj})/2}}a_{0p\alpha j}(D) s^{\alpha},
\end{eqnarray}
where $\mathrm{dim}({\mathcal A}_{0pj})$ is the mass dimension of ${\mathcal A}_{0pj}$. And the coefficient $a_{0p\alpha j}$ depends only on $D$. For fixed $\left|\alpha\right|$, the total number of terms in the $\alpha$ summation is
\begin{eqnarray}
n_\alpha = 
\left(
\begin{matrix}
\left|\alpha\right| + t - 1
\\
t -1
\end{matrix}
\right).
\end{eqnarray}
Then it can be obtained that
\begin{eqnarray}
\left|\alpha\right| - p = w_0
\end{eqnarray}
by defining 
\begin{eqnarray}
w_0 \equiv \mathrm{dim}({\mathcal C}_i)/2 - L{D}/2 + N_\nu - m_i^{\rm max}.
\end{eqnarray}
In practice ${\mathcal C}_i$ can be truncated to fixed order $p_0$, i.e.,
\begin{eqnarray}\label{Ci}
{\mathcal C}_i =&&\eta^{\mathrm{dim}({\mathcal C}_i)/2-w_0}\Bigg(\sum_{p=0}^{p_0} \sum_j\sum_{\substack{{\alpha_1,\dots,\alpha_t}\\|\alpha| = w_0+p}} \Big(a_{0p\alpha j}(D)\eta^{-p}\nonumber\\&&\times s^{\alpha} I^{(vac),D}_{L,j}\Big)+\mathcal{O}(\eta^{-p_0-1})\Bigg).
\end{eqnarray}

For the given amplitude one can choose a proper set of modified master integrals $\{\widetilde I_{j}(\eta)\}^S_{j=1}$ as shown in Eq. \eqref{modify}. Then by using Taylor series for $\eta\to\infty$ one can obtain the series representation of $\widetilde I_{k}(\eta)$,
\begin{equation}\label{e3}
\widetilde I_{k}(\eta) = \eta^{L{D}/2-N_{k}}
\sum_{p=0}^\infty \sum_j{\mathcal A}_{kpj}\eta^{-p}I^{(vac),D}_{L,j},
\end{equation}
where $N_{k}$ is the summation of the exponent of propagators for given $\widetilde I_{k}(\eta)$. And the series coefficient ${\mathcal A}_{kpj}$ can be expressed as the linear combination of monomials,
\begin{equation}\label{e4}
{\mathcal A}_{kpj} = \sum_{\substack{{\alpha_1,\dots,\alpha_t}\\|\alpha| = \mathrm{dim}({\mathcal A}_{kpj})/2}}a_{kp\alpha j}(D) s^{\alpha},
\end{equation}
where the coefficient $a_{kp\alpha j}$ depends only on $D$. Then we can obtain 
\begin{eqnarray}
|\alpha| - p = w_k
\end{eqnarray}
by defining
\begin{eqnarray}
w_k \equiv \mathrm{dim}(\widetilde I_{k})/2 - L{D}/2 + N_k.
\end{eqnarray}
Similarly as ${\mathcal C}_i$, $\widetilde I_{k}$ can be truncated to fixed order $p_0$, i.e.,
\begin{eqnarray}\label{Ik}
\widetilde I_{k} =&&\eta^{\mathrm{dim}(\widetilde I_{k})/2-w_k}\Bigg(\sum_{p=0}^{p_0} \sum_j\sum_{\substack{{\alpha_1,\dots,\alpha_t}\\|\alpha| = w_k+p}} \Big(a_{kp\alpha j}(D)\eta^{-p}\nonumber\\&&\times s^{\alpha} I^{(vac),D}_{L,j}\Big)+\mathcal{O}(\eta^{-p_0-1})\Bigg).
\end{eqnarray}

If the modified master integral set has been properly chosen, the reduction relation can be described by the linear relation between ${\mathcal C}_i$ and $\{\widetilde I_{j}\}^{S}_{j=1}$ as
\begin{equation}\label{rea}
Z_{i0}{\mathcal C}_i+ Z_{i1}\widetilde I_{1}+\cdots+Z_{iS}\widetilde I_{S}= 0,
\end{equation}
where $Z_{ik}$ is polynomial of $\eta$, independent kinematic variables $\{s_1,\dots,s_t\}$ and $D$. In the following the index $i$ will be suppressed for simplicity. Since each term in Eq. \eqref{rea} has the same mass dimension, we can define 
\begin{eqnarray}
d^\mathrm{tot}&&\equiv d_{0}+\mathrm{dim}({\mathcal C}_i) = d_{1}+\mathrm{dim}(\widetilde I_{1})\nonumber\\&&=\cdots=d_{S}+\mathrm{dim}(\widetilde I_{S}),
\end{eqnarray}
where 
\begin{eqnarray}
d_k \equiv \mathrm{dim}(Z_k)  =\left\{
\begin{aligned}
&d^\mathrm{tot} - \mathrm{dim}({\mathcal C}_i)\quad (k=0),\\
&d^\mathrm{tot} - \mathrm{dim}(\widetilde I_{k})\quad (1 \leqslant k \leqslant S).
\end{aligned}\right.
\end{eqnarray}
Therefore, $Z_{k}$ can be written as
\begin{eqnarray}\label{Zik}
Z_{k} =\sum_{\substack{\lambda_1,\dots,\lambda_t\\|\lambda|\leqslant d_{k}/2}}z_{k\lambda_0\lambda}(D)\eta^{\lambda_0}s^{\lambda},
\end{eqnarray}
where $\lambda$ is a $t$-tuple of nonnegative integers. And $\lambda_0 = d_k/2-|\lambda|$ is a nonnegative integer. The unknown coefficient $z_{k\lambda_0\lambda}$ depends only on $D$. For the expression of $Z_k$, the total number of terms in the $\lambda$ summation is 
\begin{eqnarray}
n_k  = \left(
\begin{matrix}
d_{k}/2 + t
\\
t
\end{matrix}
\right).
\end{eqnarray}

In order to obtain the explicit expressions of $\{z_{k\lambda_0\lambda}\}$, we can substitute Eqs. \eqref{Ci},\eqref{Ik} and \eqref{Zik} into Eq. \eqref{rea} and obtain 
\begin{eqnarray}\label{sum}
\eta^{d^\mathrm{tot}/2-w_\mathrm{min}}\Bigg(&&\sum_{\rho_0=0}^{p_0}\sum_{j}\sum_{\substack{\rho_1,\dots,\rho_t\\|\rho|=\rho_0+w_\mathrm{min}}}\Big(\sigma_{\rho_0\rho j}\eta^{-\rho_0}
s^{\rho}I^{(vac),D}_{L,j}\Big)\nonumber\\&&+\mathcal{O}(\eta^{-p_0-1})\Bigg) = 0,
\end{eqnarray}
where
\begin{eqnarray}
w_\mathrm{min} = \min\{w_0,\dots,w_S\},
\end{eqnarray}
and
\begin{eqnarray}\label{sigma}
\sigma_{\rho_0\rho j} = \sum_{k=0}^S\sum_{\substack{p=0\\p\geqslant|\rho|-d_k/2-w_k}}^{p_0} \, \sum_{\substack{{\alpha_1,\dots,\alpha_t}\\|\alpha| = w_k+p}}a_{kp\alpha j}z_{k\lambda_0\lambda}
\end{eqnarray}
with $\lambda = \rho - \alpha$ and $\lambda_0 = d_k/2 - |\rho| +|\alpha|$.
Since $\eta^{-\rho_0}s^{\rho}I^{(vac),D}_{L,j}$ are linear independent, their coefficients $\sigma_{\rho_0\rho j}$ should be zero. Then we obtain a system of linear equations
\begin{eqnarray}\label{equ}
\{\sigma_{\rho_0\rho j}=0\}.
\end{eqnarray}
The sets $\{\sigma_{\rho_0\rho j}\}$ and $\{z_{k\lambda_0\lambda}\}$ can be ordered by using certain well order relation, e.g. lexicographical ordering, for $(\rho_0,\rho,j)$ and $(k,\lambda_0,\lambda)$, respectively. And $\sigma_u$ and $x_v$ can be denoted as the $u$-th and $v$-th element in the corresponding set. Then Eq. \eqref{equ} can be transformed into the null space problem
\begin{eqnarray}
\begin{pmatrix}
\mathbb{M}_{1,1} & \cdots & \mathbb{M}_{1,n_c}   \\ 
\vdots   &  & \vdots  \\
\mathbb{M}_{n_e,1} & \cdots  & \mathbb{M}_{n_e,n_c}  \\
\end{pmatrix}
\begin{pmatrix}
z_1 \\\vdots \\z_{n_c}
\end{pmatrix} = 
\begin{pmatrix}
0\\\vdots\\0
\end{pmatrix},
\end{eqnarray} 
where the matrix element can be explicitly obtained by 
\begin{eqnarray}
\mathbb{M}_{uv} = \frac{\partial\sigma_u}{\partial z_v}.
\end{eqnarray}
For given $d^\mathrm{tot}$, the number of unknown coefficients $n_c = \lvert\lbrace z_{k\lambda_0\lambda}\rbrace\rvert$ is fixed while the number of equations $n_e = \lvert\lbrace\sigma_{\rho_0,\rho,j}\rbrace\rvert$ depends on the truncation order ${p_0}$. Therefore, if $d^\mathrm{tot}$ is large enough, by expanding ${\mathcal C}_i$ and $\{\widetilde I_{j}\}^{S}_{j=1}$ to higher order one can obtain enough equations ($n_e>n_c$) for the solution of the null space. 

Empirically the choice of $d^\mathrm{tot}$ can start from the minimum of the allowed values,
\begin{eqnarray}
d^\mathrm{tot}_\mathrm{min} =\max\{\mathrm{dim}({\mathcal C}_i),\min\{\mathrm{dim}(\widetilde I_{1}),\dots,\mathrm{dim}(\widetilde I_{S})\}\}.
\end{eqnarray}
If we could not find the non-trivial null space, the $d^\mathrm{tot}$ will be increased by two. Once the non-trivial null space is found, we can expand ${\mathcal C}_i$ and $\{\widetilde I_{j}\}^{S}_{j=1}$ to higher order for more equations to check the correctness and uniqueness of the solution.

Finally the modified amplitude can be written as
\begin{equation}
\widetilde {\mathcal M}(\eta) = \sum_i \sum_{k=1}^S C_{ik}(\eta) \widetilde I_{k}(\eta){\mathcal F}_i
\end{equation}
where $C_{ik} = -Z_{ik}/Z_{i0}$ is the reduction coefficient of relevant $\widetilde I_{k}(\eta)$ and ${\mathcal F}_i$ for modified
loop amplitude.
In the conventional approach the reduction coefficients could be obtained by using
tensor reduction and IBP reduction,
which could be very difficult as been reviewed in previous section.
However, as we have shown above by directly implementing series representation on
modified loop amplitude,
the difficulties in both tensor reduction and IBP reduction can be relieved.
And the final reduction relation for the original loop amplitude can be obtained by
taking the limit $\eta\to 0^+$,
\begin{equation}
{\mathcal M}
= \lim_{\eta\to 0^+} \widetilde {\mathcal M}(\eta)
= \sum_i \sum_{k=1}^S \lim_{\eta\to 0^+} C_{ik}(\eta) \widetilde I_{k}(\eta){\mathcal F}_i.
\end{equation}
Although in order to achieve loop amplitude reduction this set of master integrals
themselves
may not be convenient to evaluate analytically or numerically,
one may make further apply reduction increasingly to
the final set of master integrals that can satisfy the requirement of evaluation.

\begin{figure}[!htb]
\includegraphics[width=0.3\textwidth]{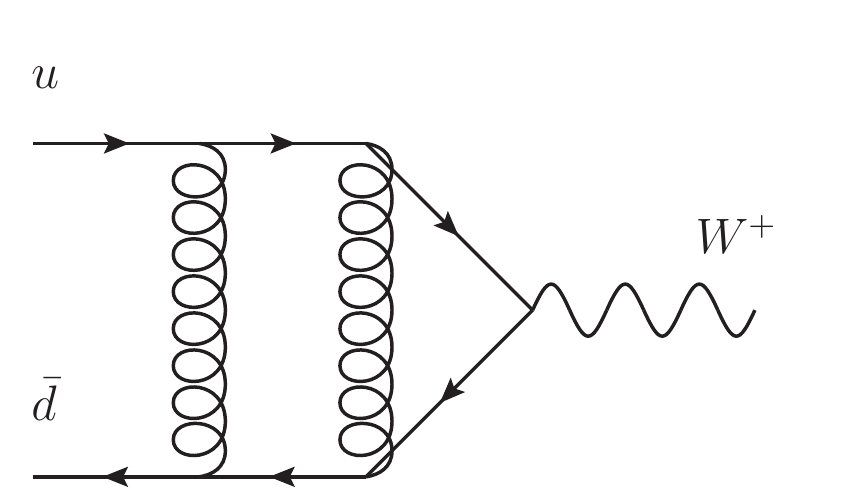}
\caption{One typical two-loop diagram for process $u\bar d\to W^+$.}
\label{WDY3_V2}
\end{figure}

\sect{Example}
In this section we take one typical two-loop diagram of W boson production shown in
Fig.\ref{WDY3_V2}
as an example to demonstrate our approach.
The diagram is plotted by using Jaxodraw
\cite{Binosi:2003yf} based on Axodraw \cite{Vermaseren:1994je}.
Its relevant modified amplitude can be written as
\begin{equation}
\widetilde {\mathcal M}(\eta) = \int \mathrm{d}^D q_1 \mathrm{d}^D q_2
\frac{N(q_1,q_2,k_1,k_2,k_3)}{
\widetilde {\mathcal D}_1
\widetilde {\mathcal D}_2
\widetilde {\mathcal D}_3
\widetilde {\mathcal D}_5
\widetilde {\mathcal D}_6
\widetilde {\mathcal D}_7
},
\end{equation}
where the denominators from loop propagators are
\begin{eqnarray}
\widetilde {\mathcal D}_1 &=& (q_1 - q_2 - k_1)^2 +\imath \eta,
\end{eqnarray}
\begin{eqnarray}
\widetilde {\mathcal D}_2 &=& (q_1 + k_2)^2 +\imath \eta,
\end{eqnarray}
\begin{eqnarray}
\widetilde {\mathcal D}_3 &=& (q_2 + k_1 + k_2)^2 +\imath \eta,
\end{eqnarray}
\begin{eqnarray}
\widetilde {\mathcal D}_5 &=& (q_1)^2 +\imath \eta,
\end{eqnarray}
\begin{eqnarray}
\widetilde {\mathcal D}_6 &=& (q_2)^2 +\imath \eta,
\end{eqnarray}
and
\begin{eqnarray}
\widetilde {\mathcal D}_7 &=& (q_1 - k_1)^2 +\imath \eta.
\end{eqnarray}
And to make complete integral family for two-loop one-final-state amplitude
we need additional one denominator
\begin{eqnarray}
\widetilde {\mathcal D}_4 &=& (q_2 + k_1)^2 +\imath \eta.
\end{eqnarray}
For reader's convenience we also explicitly show the numerator of the amplitude
\begin{eqnarray}
N(q_1,&&q_2,k_1,k_2,k_3) =\Big(\frac{\imath e g_s^4}{\sqrt{2}s_W}\Big) \frac{16}{9} \bar v(k_2) \gamma^\alpha (\not k_2 + \not
q_1 ) \gamma^\beta
\nonumber \\ && \times (\not k_3 +\not q_2 )
\not \varepsilon(k_3) P_L \not q_2 \gamma^\beta (\not k_1 - \not q_1) \gamma^\alpha
u(k_1).
\end{eqnarray}
By implementing the approach as mentioned in the previous section, we can directly extract the only form factor
\begin{equation}
{\mathcal F}_1 = \bar v(k_2) \not\varepsilon(k_3) P_L u(k_1).
\end{equation}
We can divide the Feynman parameter index set $\{1,\dots,7\}$ into three equivalence classes $[{i_1}] = \{2,5,7\}$, $[{i_2}] = \{3,4,6\}$ and $[{i_3}] = \{1\}$.
Then we can insert three unit integrals
\begin{eqnarray}
\int&& {\mathrm d} y_1 \delta(y_1-x_2-x_5-x_7)=1,\nonumber\\
\int&& {\mathrm d} y_2 \delta(y_2-x_3-x_4-x_6)=1,\nonumber\\
\int&& {\mathrm d} y_3 \delta(y_3-x_1)=1.
\end{eqnarray}
After integrating the $x_1,\dots,x_7$, the coefficient of ${\mathcal F}_1$ can be exprssed by two-loop scalar vacuum bubble integrals.
And it is known that at two-loop level there are two vacuum bubble master
integrals,
\begin{equation}
I^{(vac),D}_{2,1} \equiv \int \frac{\mathrm{d}^D q_1 \mathrm{d}^D q_2}
{\left[q_1^2+\imath\right]
\left[q_2^2+\imath\right]
\left[(q_1+q_2)^2+\imath\right]}
\end{equation}
and
\begin{equation}
I^{(vac),D}_{2,2} \equiv \int \frac{\mathrm{d}^D q_1 \mathrm{d}^D q_2}
{\left[q_1^2+\imath\right]
\left[q_2^2+\imath\right]}.
\end{equation}
In series representation the modified loop amplitude can be expressed as
\begin{widetext}
\begin{eqnarray}
\widetilde {\mathcal M}(\eta) &=&
\Big(\frac{\imath e g_s^4}{\sqrt{2}s_W}\Big)
{\mathcal F}_1
\eta^{D-4}
\Big\{
-\frac{8  (D-3) (D-2)^2 ( D^3-3 D^2+11 D-6 )}{243 D} \imath I^{(vac),D}_{2,1}
+\frac{(D-2)^4 (D^2-16 D+12)}{81 D} I^{(vac),D}_{2,2}
\nonumber \\ &&
-\frac{4 (D-3)
(5 D^7-53 D^6+319 D^5-638 D^4-1844 D^3+4552 D^2+2528 D+3456)
}{6561 D (D+2)}
\frac{m_W^2}{\eta} I^{(vac),D}_{2,1}
\nonumber \\ &&
-\frac{(D-2)^2
(83 D^6-724 D^5-976 D^4+15968 D^3-7600 D^2-51904 D-27648)
}{17496 D (D+2)}
\frac{m_W^2}{\eta} \imath I^{(vac),D}_{2,2}
+{\mathcal O}(\frac{1}{\eta^2})
\Big\}.
\end{eqnarray}
\end{widetext}

Finally for the matching procedure we choose 25 master integrals,
\begin{eqnarray}
&&
\widetilde I_1(\eta) \equiv \widetilde I_{0,1,1,0,0,1,1}(\eta),~
\quad
\widetilde I_2(\eta) \equiv \widetilde I_{0,1,1,0,1,1,1}(\eta),~
\nonumber
\end{eqnarray}
\begin{eqnarray}
&&
\widetilde I_3(\eta) \equiv \widetilde I_{0,0,1,0,1,0,1}(\eta),~
\quad
\widetilde I_4(\eta) \equiv \widetilde I_{1,0,1,0,0,0,1}(\eta),~
\nonumber
\end{eqnarray}
\begin{eqnarray}
&&
\widetilde I_5(\eta) \equiv \widetilde I_{1,0,1,0,1,0,1}(\eta),~
\quad
\widetilde I_6(\eta) \equiv \widetilde I_{1,0,1,0,1,1,0}(\eta),~
\nonumber
\end{eqnarray}
\begin{eqnarray}
&&
\widetilde I_7(\eta) \equiv \widetilde I_{1,0,1,0,1,1,1}(\eta),~
\quad
\widetilde I_8(\eta) \equiv \widetilde I_{1,0,1,0,1,2,0}(\eta),~
\nonumber
\end{eqnarray}
\begin{eqnarray}
&&
\widetilde I_9(\eta) \equiv \widetilde I_{1, 0, 2, 0, 1, 0, 1}(\eta),~
\quad
\widetilde I_{10}(\eta) \equiv \widetilde I_{0, 0, 1, 0, 1, 1, 1}(\eta),
\nonumber
\end{eqnarray}
\begin{eqnarray}
&&
\widetilde I_{11}(\eta) \equiv \widetilde I_{1, 1, 0, 0, 1, 0, 1}(\eta),
\quad
\widetilde I_{12}(\eta) \equiv \widetilde I_{1, 0, 0, 0, 1, 1, 1}(\eta),
\nonumber
\end{eqnarray}
\begin{eqnarray}
&&
\widetilde I_{13}(\eta) \equiv \widetilde I_{1, 1, 1, 0, 0, 0, 1}(\eta),
\quad
\widetilde I_{14}(\eta) \equiv \widetilde I_{1, 1, 1, 0, 0, 1, 1}(\eta),
\nonumber
\end{eqnarray}
\begin{eqnarray}
&&
\widetilde I_{15}(\eta) \equiv \widetilde I_{1, 1, 1, 0, 1, 0, 1}(\eta),
\quad
\widetilde I_{16}(\eta) \equiv \widetilde I_{1, 1, 1, 0, 1, 1, 1}(\eta),
\nonumber
\end{eqnarray}
\begin{eqnarray}
&&
\widetilde I_{17}(\eta) \equiv \widetilde I_{1, 1, 1, 0, 1, 2, 0}(\eta),
\quad
\widetilde I_{18}(\eta) \equiv \widetilde I_{1, 1, 1, 0, 2, 1, 0}(\eta),
\nonumber
\end{eqnarray}
\begin{eqnarray}
&&
\widetilde I_{19}(\eta) \equiv \widetilde I_{2, 0, 1, 0, 0, 0, 1}(\eta),
\quad
\widetilde I_{20}(\eta) \equiv \widetilde I_{2, 0, 1, 0, 1, 1, 0}(\eta),
\nonumber
\end{eqnarray}
\begin{eqnarray}
&&
\widetilde I_{21}(\eta) \equiv \widetilde I_{1, -1, 1, -1, 1, 1, 1}(\eta),
\quad
\widetilde I_{22}(\eta) \equiv \widetilde I_{1, 0, 1, -1, 1, 1, 1}(\eta),
\nonumber
\end{eqnarray}
\begin{eqnarray}
&&
\widetilde I_{23}(\eta) \equiv \widetilde I_{1, 0, 1, -2, 1, 1, 1}(\eta),
\quad
\widetilde I_{24}(\eta) \equiv \widetilde I_{1, 1, 1, -1, 1, 1, 1}(\eta),
\nonumber
\end{eqnarray}
\begin{eqnarray}
\widetilde I_{25}(\eta) \equiv \widetilde I_{1, 1, 1, -2, 1, 1, 1}(\eta),
\quad
\end{eqnarray}
where
\begin{equation}
\widetilde I_{\nu_1,\nu_2,\nu_3,\nu_4,\nu_5,\nu_6,\nu_7}(\eta) \equiv
\int
\frac{
\mathrm{d}^D q_1 \mathrm{d}^D q_2
}{
\widetilde {\mathcal D}_1^{\nu_1}
\widetilde {\mathcal D}_2^{\nu_2}
\widetilde {\mathcal D}_3^{\nu_3}
\widetilde {\mathcal D}_4^{\nu_4}
\widetilde {\mathcal D}_5^{\nu_5}
\widetilde {\mathcal D}_6^{\nu_6}
\widetilde {\mathcal D}_7^{\nu_7}
}.
\end{equation}
For simplicity, the factor $(\imath e g_s^4)/(\sqrt{2}s_W)$ is omitted in results. Then the coefficients between modified amplitude and master integrals are
\begin{widetext}
\begin{eqnarray}
C_{1}(\eta) =&& -\frac{16 (D-6)^2 (D-3)}{9 (D-2)}, \nonumber\\
C_{2}(\eta) =&& \frac{8 \left(\left(3 D^2-32 D+68\right) m_W^2-2 \imath (D-2)^2 \eta \right)}{9 (D-2)},\nonumber\\
C_{3}(\eta) =&& -\frac{8 \left(D^4-20 D^3+156 D^2-532 D+648\right)}{9 (D-3) D m_W^2},\nonumber\\
C_{4}(\eta) =&& \frac{8 \left(\left(2 D^5-20 D^4-23 D^3+919 D^2-3466 D+3888\right) m_W^2-\imath (3 D-8) \left(D^4-29 D^3+230 D^2-782 D+972\right) \eta \right)}{9 (D-3) (D-2) D m_W^4},\nonumber\\
C_{5}(\eta) =&&-\frac{8 \left(\left(D^4-8 D^3+16 D^2+24 D-80\right) D\, m_W^2+\imath \left(D^5-15 D^4+126 D^3-576 D^2+1352 D-1296\right) \eta \right)}{9 (D-3) (D-2) D m_W^2},\nonumber\\
C_{6}(\eta) =&& \frac{8 \left((D-4) (D-2) (2 D^3-39 D^2+204 D-300) m_W^2+2 \imath \left(18 D^5-379 D^4+3121 D^3-12452 D^2+24084 D-18128\right) \eta \right)}{9 (D-4) (D-2) (3 D-8) m_W^2},\nonumber\\
C_{7}(\eta) =&& \frac{128 (D-3) (D-2) m_W^2+8 \imath \left(2 D^4-27 D^3+162 D^2-476 D+536\right) \eta }{9 (D-3) (D-2)},\nonumber\\
C_{8}(\eta) =&& -\frac{16 \left(7 D^4-124 D^3+806 D^2-2220 D+2192\right) \eta  \left(4 \eta -\imath m_W^2\right)}{9 (D-4) (D-2) (3 D-8) m_W^2},\nonumber\\
C_{9}(\eta) =&& \frac{4 \left(4 \left(D^3-8 D^2+38 D-68\right) \eta ^2-2 \imath \left(3 D^3-45 D^2+218 D-328\right) \eta \, m_W^2+(D-6) (D+2) m_W^4\right)}{9 (D-3) (D-2) m_W^2},\nonumber\\
C_{10}(\eta) =&& \frac{8 \left(\left(D^3-15 D^2+80 D-147\right) (3 D-8) (D-2) m_W^2+2 \imath \left(2 D^4-34 D^3+219 D^2-610 D+616\right) \eta \right)}{9 (D-3) (D-2) (3 D-8) m_W^2},\nonumber\\
C_{11}(\eta) =&& \frac{16 (D-2) \left(m_W^2+2 \imath \eta \right)}{9 m_W^2},\nonumber\\
C_{12}(\eta) =&& \frac{8}{9 (D-4) (D-3)^2 D (3 D-8) m_W^4}\big\{3 (3 D-8) (D-4) \left(D^4-29 D^3+230 D^2-782 D+972\right) \eta ^2\nonumber\\
&&+\imath D\left(15 D^5-326 D^4+2797 D^3-11752 D^2+24052 D-19152\right)\eta\,m_W^2-16 (D-4) (D-3)^2 D (3 D-8) m_W^4\},\nonumber\\
C_{13}(\eta) =&& \frac{8 \left((D-4) \left(D^3-19 D^2+128 D-236\right) m_W^2-4 \imath \left(D^4-19 D^3+126 D^2-348 D+344\right) \eta \right)}{9 (D-4) (D-2) m_W^2},\nonumber\\
C_{14}(\eta) =&& \frac{8 \left(\left(3 D^2-32 D+68\right) m_W^2+4 \imath \left(D^2-12 D+28\right) \eta \right)}{9 (D-2)}, \nonumber\\
C_{15}(\eta) =&& \frac{64 \left(2 (D-2) m_W^2+\imath \left(2 D^2-15 D+30\right) \eta \right)}{9 (D-2)},\nonumber\\
C_{16}(\eta) =&& \frac{16 \left(-4 \imath \left(D^2-6 D+10\right) \eta\,  m_W^2+(D-5) (D-2)^2 \eta ^2-4 (D-2) m_W^4\right)}{9 (D-2)},\nonumber\\
C_{17}(\eta) =&& \frac{4 (D-6) \left(-m_W^2-4 \imath \eta \right) \left((D+2) m_W^2-2 \imath (D-6) (D-3) \eta \right)}{9 (D-3) (D-2)},\nonumber\\
C_{18}(\eta) =&& \frac{32 \left(D^3-14 D^2+68 D-104\right) \eta ^2}{9 (D-4) (D-3) (D-2)},\nonumber\\
C_{19}(\eta) =&& \frac{8}{9 (D-3) (D-2) D m_W^4}\big\{\left(6 D^3-89 D^2+446 D-648\right) m_W^4+6 \left(D^4-29 D^3+230 D^2-782 D+972\right) \eta ^2\nonumber\\&&-\imath \left(4 D^4-137 D^3+1259 D^2-4826 D+6480\right) \eta \,  m_W^2\},\nonumber\\
C_{20}(\eta) =&& \frac{4}{9 (D-4) (D-3) (D-2) (3 D-8) m_W^2}\big\{2 \imath (D-4) \left(D^4-24 D^3+156 D^2-280 D+48\right) \eta \, m_W^2\nonumber\\&&
+4 \left(-22 D^5+454 D^4-3647 D^3+14094 D^2-26240 D+18912\right) \eta ^2+(3 D-8) (D-6) (D-4) (D+2) m_W^4\},\nonumber\\
C_{21}(\eta) =&& \frac{8 \left(D^3-14 D^2+68 D-104\right)}{9 (D-2) m_W^2}, \nonumber\\
C_{22}(\eta) =&& -\frac{8 \left((D-6) \left(D^2-8 D+20\right) m_W^2+2 \imath (D-2)^2 \eta \right)}{9 (D-2) m_W^2}, \nonumber\\
C_{23}(\eta) =&& \frac{16 (D-2)}{9 m_W^2},\nonumber\\
C_{24}(\eta) =&& \frac{16 \imath \left(\left(D^3-16 D^2+80 D-136\right) \eta +8 \imath m_W^2\right)}{9 (D-2)},\nonumber\\
C_{25}(\eta) =&& -\frac{16 (D-2) \left(m_W^2+2 \imath \eta \right)}{9 m_W^2}.
\end{eqnarray}
\end{widetext}
By checking the asymptotic behavior of above master integrals at $\eta\rightarrow
0^+$,
we found that master integral $\widetilde I_{3}(\eta), \widetilde I_{10}(\eta),
\widetilde I_{11}(\eta), \widetilde I_{12}(\eta)$, vanish.
Also some of the relevant coefficients of the master integrals,
$C_{8}(\eta)$ and $C_{18}(\eta)$,
become zero in the limit.
Then finally we found 19 non-vanishing master integrals and their relevant coefficients, the limits of remaining non-vanishing coefficients are
\begin{eqnarray}
	&&C_1(0)=-\frac{16 (D-6)^2 (D-3)}{9 (D-2)},
	\nonumber\\
	&&C_2(0)=C_{14}(0)=\frac{8 (3 D^2-32 D+68) m_W^2}{9 (D-2)},
	\nonumber\\
	&&C_4(0)=\frac{8 (2 D^5-20 D^4-23 D^3+919 D^2-3466 D+3888)}{9 (D-3)
		(D-2) D m_W^2},
	\nonumber\\
	&&C_5(0)=-\frac{8 (D^4-8 D^3+16 D^2+24 D-80)}{9 (D-3) (D-2)},
	\nonumber\\
	&&C_6(0)=\frac{8 (2 D^3-39 D^2+204 D-300)}{9 (3 D-8)},
	\nonumber\\
	&&C_7(0)=C_{15}(0)=\frac{128 m_W^2}{9},
	\nonumber\\
	&&C_{9}(0)=C_{20}(0)=\frac{4 (D-6) (D+2) m_W^2}{9 (D-3) (D-2)},
	\nonumber\\
	&&C_{13}(0)=\frac{8 (D^3-19 D^2+128 D-236)}{9 (D-2)},
	\nonumber\\
	&&C_{16}(0)=-\frac{64 m_W^4}{9},
	\nonumber\\
	&&C_{17}(0)=-\frac{4 (D-6) (D+2) m_W^4}{9 (D-3) (D-2)},
	\nonumber\\
	&&C_{19}(0)=\frac{8 (6 D^3-89 D^2+446 D-648)}{9 (D-3) (D-2) D},
	\nonumber\\
	&&C_{21}(0)=\frac{8 (D^3-14 D^2+68 D-104)}{9 (D-2) m_W^2},
	\nonumber\\
	&&C_{22}(0)=-\frac{8 (D-6) (D^2-8 D+20)}{9 (D-2)},
	\nonumber\\
	&&C_{23}(0)=\frac{16 (D-2)}{9 m_W^2},
	\nonumber\\
	&&C_{24}(0)=-\frac{128 m_W^2}{9(D-2)},
	\nonumber\\
	&&C_{25}(0)=-\frac{16 (D-2)}{9}.
\end{eqnarray}
The explicit expressions of the coefficients are consistent
with the results in the conventional approach using FeynCalc \cite{Shtabovenko:2016sxi} and FIRE5 \cite{Smirnov:2014hma}.

\sect{Conclusions}
In this paper, based on series representation
we propose an alternative reduction approach to directly reduce loop amplitude
into linear combination of master integrals and extract the form factors meanwhile.
This approach can relieve the difficulties in tensor reduction and IBP reduction
for complicated scattering processes.
This approach has been demonstrated in one typical two-loop Feynman diagram for
the W boson production.

\sect{Acknowledgments}
This work was supported by the National Natural Science Foundation of China under
Grant No. 11675185.
The authors want to thank Yan-Qing Ma, Xiao Liu, Yang Zhang, Xiao-Hui Liu, Yu Jia and Hao Zhang for helpful
discussions. Najam ul Basat would like to acknowledge financial support from
CAS-TWAS President's Fellowship Program 2017.

\bibliography{reduce}
\end{document}